\documentclass[twocolumn,showpacs,preprintnumbers,amsmath,amssymb,superscriptaddress]{revtex4}
\usepackage[maccyr]{inputenc}
\usepackage[T2A]{fontenc}
\usepackage[english,russian]{babel}
\usepackage{bm}
\usepackage{epsfig,amssymb,amsmath,bm}

\begin{document}

\title{The first order equations for scalar bosons}
\author{K. S. Karplyuk}
\email{karpks@hotmail.com}
 \affiliation{Department of Radiophysics, Taras Shevchenko University, Academic
Glushkov prospect 2, building 5, Kyiv 03122, Ukraine}
\author{O. O. Zhmudskyy}\email{ozhmudsky@physics.ucf.edu}
 \affiliation{Department of Physics, University of Central Florida, 4000 Central Florida Blvd. Orlando, FL, 32816 Phone: (407)-823-4192}
\begin{abstract}
It is shown that, in contrast to the generally accepted opinion, there exist first order equations for the scalar bosons.  
Such equations are proposed below.   They are similar to the Proca equations and Maxwell equations for the vector bosons.    
\end{abstract}

\pacs{03.65.Pm}

\maketitle
There exists a generally accepted opinion that scalar bosons (with spin 0) can not be described by the first 
order equations (see for example \cite{ryd}, \S 2.8).  These bosons are descried by a scalar function, which satisfies the 
second order Klein-Gordon equation  \cite{s}-\cite{p}:   
\begin{equation}
\Box S+\varkappa^2 S=0.
\end{equation}
Here $\square=\partial_\alpha\partial^\alpha=c^{-2}\partial^2/\partial t^2-\Delta$ is the d'Alambetian, $\varkappa=mc/\hbar$. 

It is shown below that this opinion is not correct. We can use the following analogy.  
 Vector bosons are described by s vector potential, the components of which  satisfy  Klein-Gordon equation.  
 It is natural to assume that scalar bosons must be  described by a scalar potential, which satisfies  the same equation. 
 For scalar bosons it is natural to treat  $S$ in (1) as a potential.  
 Field equations for massless vector bosons (Maxwell's equations) connect 2-nd rank tensor field components  (differs from potential 
  dimension).   
 Field equations for massive vector bosons (Proca equations) connect 2-nd rank field components (non-potential dimension) and vector potential.    
 
 It is natural to connect the vector field $V_\alpha$  (differs from the $S$ dimension) and field $\epsilon$ (the same dimension as for the $S$) 
 with massive vector bosons.  
 
\begin{equation}
V_\alpha=\partial_\alpha S,\hspace{7mm}\epsilon=\varkappa S.
\end{equation}
Then the usual Lagrangian density for scalar bosons can be written as:    
\begin{equation}
L=\underbrace{\frac{1}{2}(V_\alpha V^\alpha-\epsilon^2)}_{L_0}+\underbrace{\zeta sS}_{L_i}.
\end{equation}

Here $L_0$ is the Lagrangian density of the free scalar field.  For generalization we also include in $L$ a scalar source density $s$, 
which creates a scalar field and some coefficient  $\zeta$, which depends on the choice of the system of units.     
Interaction of the scalar field with the source is given by the Lagrangian density $L_i$, which is the analogy to the 
term $j_\mu A^\mu$ in electrodynamics.  An equation for Euler's density can be obtained from (3) if the potential is used as an 
independent variable.   

\begin{equation}
\partial_\alpha V^\alpha+\varkappa \epsilon=\zeta s.
\end{equation}
Equation (4) is the first order equation for the scalar bosons.  In a three dimensional representation it can be written as:  
\begin{equation}
\frac{1}{c}\frac{\partial V_0}{\partial t}+\mathrm{div}\bm{V}+\varkappa \epsilon=\zeta s.
\end{equation}
Equation (4) has the same structure as a corresponding equation for vector bosons:  
\[\partial_\alpha F^{\alpha\beta}+\varkappa^2 A^\beta=\zeta j^\beta.\]  
Both equations connect the 4-divergence of the vector or tensor field with the scalar or vector source.   
Besides, the following equations are satisfied due to the definitions (2):  
\begin{gather}
\partial_\alpha V_\beta-\partial_\beta V_\alpha=0,\\
\partial_\alpha\epsilon-\varkappa V_\alpha=0.
\end{gather}
In three dimensional representations these equations look like:
\begin{gather}
\frac{1}{c}\frac{\partial\bm{V}}{\partial t}+\nabla V_0=0,\\
\nabla\times \bm{V}=0,\\
\frac{1}{c}\frac{\partial\epsilon}{\partial t}-\varkappa  V_0=0,\\
\nabla\epsilon+\varkappa \bm{V}=0.
\end{gather}
Equations (4)-(11) are the field equations of first order for massive scalar bosons.  Equations (4)-(5) are an analogy of the 
nonuniform Proca equation.  Equations (6)-(11) are an analogy of the uniform Proca equations.  

It is evident that for the massive bosons equation (6) follows from equation (7) and can be omitted. For massless bosons the 
field $\epsilon$  is zero due to the definition (2).  That is why we must keep equation (6).    
Equations (4)-(11) for the massless boson become:    
\begin{gather}
\frac{1}{c}\frac{\partial V_0}{\partial t}+\mathrm{div}\bm{V}=\zeta s,\\
\frac{1}{c}\frac{\partial\bm{V}}{\partial t}+\nabla V_0=0,\\
\nabla\times \bm{V}=0.
\end{gather}
Equation (12) is an analogy of the nonuniform Maxwell equations, and equations (13)-(14) are an analogy of uniform equations.  
Let us express fields versus potentials and substitute them into (4).  It gives us an equation for the potential:  
\begin{equation}
\Box S+\varkappa^2 S=\zeta s.
\end{equation}

In previous discussion we dealt with uncharged scalar bosons.  Description of charged bosons requires complex fields.  
In this case the usual Lagrangian density for charged scalar bosons is:   
\begin{equation}
L=(V_\alpha V^{\alpha *}-\epsilon\epsilon^*)+\zeta(sS^*+s^*S).
\end{equation}  
We can rederive the same field equations (4)-(11)  for charged bosons ( and their complex-conjugate) from (16) by using 
$S$ and $S^*$ as independent variables.  

Conservation laws for scalar bosons can be obtained by combining equations (4)-(11) and their complex-conjugate in the same way 
as we obtain conservation laws in electrodynamics by combining Maxwell's equations.  So, we can derive an energy conservation law:    
\begin{gather}
\frac{1}{c}\frac{\partial}{\partial t}(V_0V_0^*+\bm{V}\bm{V}^*+\epsilon\epsilon^*)+\mathrm{div}(V_0\bm{V}^*+V_0^*\bm{V})=\nonumber\\=\zeta(V_0s^*+V_0^*s),
\end{gather}
A linear momentum conservation law: 
\begin{gather}
\frac{1}{c}\frac{\partial}{\partial t}(V_0\bm{V}^*+V_0^*\bm{V})+\bm{V}\mathrm{div}\bm{V}^*+\bm{V}^*\mathrm{div}\bm{V}+\nonumber\\+
\mathrm{grad}((V_0V_0^*-\epsilon\epsilon^*)=\zeta(\bm{V}s^*+\bm{V}^*s)
\end{gather}
and a charge conservation law:  
\begin{gather}
\frac{1}{c}\frac{\partial}{\partial t}\,i(\epsilon V_0^*-\epsilon^*V_0)+\mathrm{div}\,i(\epsilon\bm{V}^*-\epsilon^*\bm{V})=\zeta i(\epsilon s^*-\epsilon^*s).
\end{gather}  
The left-hand sides of equations (17)-(18) are the 4-divergence of the energy-impulse tensor of the fields  $V_\alpha$ and $\epsilon$,  
which are connected with scalar bosons.  Also, this tensor can be obtained in a usual way from the Lagrangian density $L_0$:   
\begin{gather}
T^{\mu\nu}=\frac{\partial L_0}{\partial(\frac{\partial S}{\partial x^\mu})}\frac{\partial S}{\partial x_\nu}+
\frac{\partial L_0}{\partial(\frac{\partial S^*}{\partial x^\mu})}\frac{\partial S^*}{\partial x_\nu}-\eta^{\mu\nu}L_0=\nonumber\\
=V^\mu V^{\nu*}+V^\nu V^{\mu*}-\eta^{\mu\nu}(V_\alpha V^{\alpha *}-\epsilon\epsilon^*).
\end{gather}
The divergence $\partial T^{\mu\nu}/\partial x^\mu$ coincides with the left-hand sides of  equations (17)-(18).   The energy density is a positively  
defined function:  
\begin{equation}
T^{00}=V_0V_0^*+\bm{V}\bm{V}^*+\epsilon\epsilon^*.
\end{equation}
The left-hand sides of  equation (19) is the  4-divergence of the Noether's current.  This current is connected  with the invarince of the Lagrangian density    
$L_0 = V_\alpha V^{\alpha *}-\epsilon\epsilon^*$  
with respect to the global transformations $S\to e^{i\alpha}S$ of the group $U(1)$.  In order to  ensure  the invariance with 
respect to the local transformations of the group $U(1)$, the substitution $\partial_\mu\to\partial_\mu+iqA_\mu/\hbar c$ must be done in equations  (16) and (4)-(11).   In this way we can obtain first order equations which describe the interaction of charged scalar bosons with an electromagnetic field:  
\begin{gather}
(\partial_\alpha+i\frac{q}{\hbar c}A_\alpha) V^\alpha+\varkappa \epsilon=\zeta s,\\
(\partial_\alpha+i\frac{q}{\hbar c}A_\alpha)V_\beta-(\partial_\beta+i\frac{q}{\hbar c}A_\beta) V_\alpha=0,\\
(\partial_\alpha+i\frac{q}{\hbar c}A_\alpha)\epsilon-\varkappa V_\alpha=0.
\end{gather}
In three dimensional representations these equations look like:
\begin{gather}
\frac{1}{c}\frac{\partial V_0}{\partial t}+\mathrm{div}\bm{V}+\varkappa \epsilon+i\frac{q}{\hbar c}A_0V_0-i\frac{q}{\hbar c}\bm{A}\cdot \bm{V}=\zeta s,\\
\frac{1}{c}\frac{\partial\bm{V}}{\partial t}+\nabla V_0+i\frac{q}{\hbar c}A_0\bm{V}-i\frac{q}{\hbar c}\bm{A}V_0=0,\\
\nabla\times \bm{V}-i\frac{q}{\hbar c}\bm{A}\times\bm{V}=0,\\
\frac{1}{c}\frac{\partial\epsilon}{\partial t}-\varkappa  V_0+i\frac{q}{\hbar c}A_0\epsilon=0,\\
\nabla\epsilon+\varkappa \bm{V}-i\frac{q}{\hbar c}\bm{A}\epsilon=0.
\end{gather}  
In the particular case of a scalar boson with charge $q=-e$ in a Coulomb's field $A_0=Ze/r$, $\bm{A}=0$ we  find using (25)-(29): 
\begin{equation}
\left(\frac{1}{c}\frac{\partial}{\partial t}-i\frac{Z\alpha}{r}\right)^2\epsilon-\Delta\epsilon+\varkappa^2\epsilon=0.
\end{equation}
Here $\alpha=e^2/\hbar c$ and $s=0$.  This equation coincides with the one initially proposed by Schrodinger for taking into account the  
contribution of relativistic effects in a hydrogen spectrum \cite{s}.  It is well-known that the spectrum obtained from (30) is not in good agreement with  experimental results \cite{sch}.    

 Quantization of the equations (4)-(11) coincides with the results obtained in \cite{p} for Klein-Gordon equation. 

Our above discussion shows that bosons with spin 0 can be described by the first order equations as well as for bosons with spin 1.

\end{document}